\documentstyle[11pt]{article} 
\begin{document} \begin{titlepage} 
\begin{flushright}
CfPA 97-th-26\\ 
astro-ph/9712121\\ 
(Submitted to {\bf Physical Review})\\ 
\end{flushright} 
\begin{center} 
\Large {\bf Power Spectrum Estimators For Large CMB Datasets}\\ 
\vspace{1cm} 
\normalsize
\large{Julian Borrill}\\ 
\normalsize 
\vspace{.5cm} 
Center for Particle Astrophysics, University of California, Berkeley, CA 94720\\ 
and\\
National Energy Research Scientific Computing Center, Lawrence
Berkeley National Laboratory, University of California, Berkeley, CA
94720\\ 
\vspace{.5cm} 
\end{center} 
\baselineskip=24pt 
\begin{abstract}
Forthcoming high-resolution observations of the Cosmic Microwave
Background (CMB) radiation will generate datasets many orders of
magnitude larger than have been obtained to date. The size and
complexity of such datasets presents a very serious challenge to
analysing them with existing or anticipated computers. Here we present
an investigation of the currently favored algorithm for obtaining the
power spectrum from a sky-temperature map --- the quadratic
estimator. We show that, whilst improving on direct evaluation of the
likelihood function, current implementations still inherently scale as
the equivalent of $O(N_{p}^{3})$ in the number of pixels or worse, and
demonstrate the critical importance of choosing the right
implementation for a particular dataset.
\end{abstract}
\begin{center} 
\end{center} 
\end{titlepage}


\section{Introduction}

Over the next ten years a number of ground-based, balloon-borne and
satellite observations of the Cosmic Microwave Background (CMB) are
planned with sufficient resolution to determine the CMB power spectrum
up to multipoles $l \sim 1000$ or more (for a general review of
forthcoming observations see \cite{S}). According to current theory
this will provide us with the locations, amplitudes, and shapes of the
Doppler peaks, and hence the values of the fundamental cosmological
parameters to unprecedented accuracy. The CMB will then have lived up
to its promise of being the most powerful discriminant between
cosmological models \cite{Kn,HSS,KKJS}.

In preparation for these datasets considerable effort is being put
into developing ways of extracting the information they contain.
Typically the raw data is cleaned and converted into a time-ordered
dataset. This is then turned into a sky temperature map, and the map
analysed to find its power spectrum. Having obtained the power
spectrum of the dataset we can compare it with the predictions of any
class of cosmological models to determine the most likely values of
the parameters associated with that class. Whilst it would also be
possible to estimate such cosmological parameters directly from the
data, this would require the assumption of a class of models during
the data analysis. We therefore choose to provide the more generic
result of the power spectrum.

Here we consider the analysis of an $N_{p}$ pixel map from a simple
pointing experiment for multipoles $1 \leq l \leq N_{l}$ in bins $1
\leq b \leq N_{b}$ --- ie. we determine the location of and the
curvature about the peak of the maximum likelihood function of the
binned power spectrum coefficients $C_{b}$.

\section{Maximum Likelihood Analysis}

Any observation of the CMB contains both signal and noise
\begin{equation}
\Delta_{i} = s_{i} + n_{i}
\end{equation}
at each pixel. For independent, zero-mean, signal and noise the
covariance matrix of the data
\begin{equation}
M \equiv \left< \Delta \, \Delta^{T} \right> = \left< s \, s^{T}
\right> + \left< n \, n^{T} \right>
\end{equation}
is a symmetric, positive definite and dense. Given any binned power
spectrum $C_{b}$ and a shape parameter $C_{l}^{s}$ within each bin
such that
\begin{equation}
C_{l} = C_{l}^{s} C_{b} \;\;\;\;\;\; l \in b
\end{equation}
we can construct the signal covariance matrix; for a simple pointing
experiment this is
\begin{eqnarray}
\label{eq.scm}
S_{i i'} \equiv \left< s_{i} \, s_{i'} \right> & = &
\sum_{l=0}^{N_{l}} \, \frac{2 l + 1}{4 \pi} \, C_{l} \, B_{l}^{2} \,
P_{l}(\cos \theta_{i i'}) \nonumber \\ [0.1in] & = &
\sum_{b=0}^{N_{b}} C_{b} \sum_{l \in b} \frac{2 l + 1}{4 \pi} \,
C_{l}^{s} \, B_{l}^{2} \, P_{l}(\cos \theta_{i i'})
\end{eqnarray}
where $B_{l}$ is the multipole beam map and $\theta_{i i'}$ is the
angular separation of pixels $i$, $i'$. Taking the CMB fluctuations to
be Gaussian is not only consistent with the favoured inflationary
cosmologies but also has the maximum entropy if we make no assumption
about the higher moments of the data predicted to be non-zero in
defect-based models. The probability of the observed dataset given the
assumed power spectrum is then
\begin{equation}
\label{eq.lf}
{\cal L}(C) \equiv P(\Delta \, | \, C) = \frac{e^{-{\small\frac{1}{2}}
\, \Delta^{T} \, M^{-1} \, \Delta}} {(2\pi)^{N_{p}/2} \left| M
\right|^{1/2}}
\end{equation}
Assuming a uniform prior, so that $P(C \, | \, \Delta) \propto
P(\Delta \, | \, C)$, the most likely power spectrum will be that
which maximizes ${\cal L}(C)$, with covariance matrix
\begin{equation}
\left[ {\cal Q}^{-1} \right]_{b b'} \equiv - \left. \frac{\partial^{2}
{\cal L}} {\partial C_{b} \, \partial C_{b'}}\right|_{C = C_{max}}
\end{equation}

\section{Direct Evaluation}

Historically the most likely power spectrum has usually been obtained
by evaluating ${\cal L}(C)$ directly over the bin parameter space to
locate its maximum (for example for the COBE data \cite{G1,G2}, with
the additional refinement of using a complete set of cut-sky basis
functions in place of the incomplete spherical harmonics). To date the
fastest general solution uses a Cholesky decomposition of the matrix
$M$, costing $O(N_{p}^2)$ in size and $O(N_{p}^{3})$ in time for a
single point in parameter space.

Algorithms for searching the $N_{b}$-dimensional parameter space ---
such as maximum gradient ascent --- typically require $O(N_{b})$
evaluations at each of many steps. Moreover, calculating the
covariance matrix at the maximum by discrete differencing requires a
further $O(N_{b}^{2})$ evaluations. Overall therefore current
implementations of this algorithm scale as $O(N_{p}^2)$ in size and
$O(N_{b}^{2} N_{p}^{3})$ in time, and become hopelessly intractable
for any of the anticipated datasets.

Although there have been some attempts to improve on this scaling ---
for example by transforming to the signal-to-noise eigenbasis
\cite{BJK}, using approximations for the determinant \cite{Ko}, or
assuming azimuthally symmetric noise \cite{OSH} --- none has provided
a fast way to search a high dimensional multipole bin parameter space
under an arbitrarily complex dataset.

\section{Quadratic Estimators}

Since we are interested both in a rapid search for the maximum of
${\cal L}$, and in evaluating the curvature matrix of ${\cal L}$ at
this maximum, we solve
\begin{equation}
\frac{\partial \ln {\cal L}}{\partial C} = 0
\end{equation}
iteratively by the Newton-Raphson method. Starting from some
(sufficiently good) target power spectrum $C$ the correction
\begin{equation}
\label{eq.nr}
\delta C = - \left[ \frac{\partial^{2} \ln {\cal L}}{\partial C_{2}}
\right]^{-1} \frac{\partial \ln {\cal L}}{\partial C}
\end{equation}
gives rapid convergence to the maximum of ${\cal L}$.

Taking the log and repeatedly differentiating equation (\ref{eq.lf})
\begin{eqnarray}
\ln {\cal L} & = & - \frac{1}{2} \left( \Delta^{T} \, M^{-1} \, \Delta
+ {\rm Tr} \left[\ln M \right] + N_{p} \ln 2 \pi \right) \nonumber \\ [0.1in]
\frac{\partial \ln {\cal L}}{\partial C_{b}} & = & 
\frac{1}{2} \left( \Delta^{T} \, M^{-1} \, \frac{\partial S}{\partial
C_{b}} \, M^{-1} \, \Delta - {\rm Tr} \left[ M^{-1} \, \frac{\partial
S}{\partial C_{b}} \right] \right) \nonumber \\ [0.1in]
\frac{\partial^{2} \ln {\cal L}}{\partial C_{b} \, \partial C_{b'}} & = & 
\frac{1}{2} \left( \Delta^{T} \left[ M^{-1} \, \frac{\partial^{2}
S}{\partial C_{b} \, \partial C_{b'}} \, M^{-1} - 2 \, M^{-1} \,
\frac{\partial S}{\partial C_{b}} \, M^{-1} \, \frac{\partial
S}{\partial C_{b'}} \, M^{-1} \right] \Delta \right. \nonumber \\
& & \;\;\;\: - \left. {\rm Tr} \left[ M^{-1} \, \frac{\partial^{2}
S}{\partial C_{b} \, \partial C_{b'}} \, M^{-1} - M^{-1} \,
\frac{\partial S}{\partial C_{b}} \, M^{-1} \, \frac{\partial S}
{\partial C_{b'}} \right] \, \right)
\end{eqnarray}
Now if instead of the computationally intensive full curvature matrix
we settle for its much simpler ensemble average (ie. the Fisher
information matrix) we have
\begin{equation}
\label{eq.fm}
F_{b b'} = - \left\langle \frac{\partial^{2} \ln {\cal L}}{\partial
C_{b} \, \partial C_{b'}} \right\rangle = \frac{1}{2} \, {\rm Tr}
\left[ M^{-1} \, \frac{\partial S}{\partial C_{b}} \, M^{-1} \,
\frac{\partial S}{\partial C_{b'}} \right]
\end{equation}
and equation (\ref{eq.nr}) reduces to
\begin{equation}
\delta C = F^{-1} \frac{\partial \ln {\cal L}}{\partial C}
\end{equation}
Note that this procedure both locates the maximum and generates the
(albeit approximated) covariance matrix $F^{-1}$.

The most computationally expensive calculation here is still the
evaluation of the Fisher matrix, for which two methods have been
proposed. Noting that, from equation (\ref{eq.scm}), the derivative
matrix for each bin
\begin{equation}
\frac{\partial S}{\partial C_{b}} = \sum_{l \in b} \frac{2 l + 1}{4
\pi} \, C_{l}^{s} \, B_{l}^{2} \, P_{l}
\end{equation}
is independent of iterative step, Bond, Jaffe and Knox \cite{BJK}
calculate them explicitly and solve 
\begin{equation}
M X_{b} = \frac{\partial S}{\partial C_{b}}
\end{equation}
column by column for each bin. The first two rows of table 1 shows the
cost of evaluating the Fisher matrix this way.

Alternatively, Tegmark \cite{T} has pointed out that each $(N_{p}
\times N_{p})$ Legendre polynomial matrix can be factorised into the
product of the corresponding $(N_{p} \times (2 l + 1))$ spherical
harmonic matrix and its transpose
\begin{equation}
\frac{2 l + 1}{4 \pi} \, P_{l} = Y_{l} \, Y_{l}^{T}
\end{equation}
where
\begin{equation}
\left[ Y_{l} \right]_{i m} = Y_{l m}(\theta_{i}, \psi_{i})
\end{equation}
for the real spherical harmonic $Y_{l m}$ in the direction of pixel
$i$. Now
\begin{equation}
\frac{\partial S}{\partial C_{b}} = 
\sum_{l \in b} C_{l}^{s} \, B_{l}^{2} \, Y_{l} \, Y_{l}^{T}
\end{equation}
and we can use the invariance of the trace of a product of matrices
under cyclic permutations to rewrite equation (\ref{eq.fm}) as
\begin{equation}
F_{b b'} = \frac{1}{2} \, \sum_{l \in b} \, \sum_{l' \in b'} \,
C_{l}^{s} \, C_{l'}^{s} \, B_{l}^{2} \, B_{l'}^{2} \, {\rm Tr} \left[
\, \left( Y_{l'}^{T} \, M^{-1} \, Y_{l} \right) \, \left( Y_{l'}^{T} 
\, M^{-1} \, Y_{l} \right)^{T} \right]
\end{equation}
and solve 
\begin{equation}
M X_{l} = Y_{l} 
\end{equation}
column by column for each multipole, and
\begin{equation}
Z_{l \, l'} = Y_{l}^{T} \, X_{l'}
\end{equation}
for each pair of multipoles, and hence each pair of bins. The last
three rows of table 1 shows the cost of evaluating the Fisher matrix
this way.

For CMB observations we have $N_{b} \ll N_{p}$, so that the first
algorithm (A1) scales as $O(N_{b} \, N_{p}^2)$ in size and $O(N_{b} \,
N_{p}^3)$ in time. Similarly $N_{l}^{2} \geq N_{p}$, with approximate
equality for all-sky maps, so that the second algorithm (A2) scales as
$O(N_{l}^{4})$ in size and $O(N_{l}^{4} \, N_{p})$ in time. Table 2
shows the implications for a range of future experiments, scaled from
implementations of each algorithm applied to an unbinned reduced COBE
dataset. Note that no assumption has been made about binning in the
MAP and PLANCK datasets.

\section{Conclusions}

We have implemented two algorithms using the quadratic estimator as a
means of determining the maximum likelihood power spectrum and its
covariance matrix from a pixelized map of the CMB. Despite previous
claims, whilst each is an improvement on direct evaluation of the
likelihood function, neither scales better in time than $O(N_{p}^{3})$
in the number of pixels in the map. Ultimately the advantage of each
is in a reduction of the scaling prefactor as compared with direct
evaluation.

Comparing the two algorithms it is apparent that the choice of which
to use for a particular dataset is critical --- with timings differing
by up to a factor of 1000. Broadly speaking, observations of small
patches of the sky, where $N_{l} \gg \surd N_{p}$, should be analysed
using A1, whilst all-sky maps, with $N_{l} \sim \surd N_{p}$, should
be analysed using A2.

All timings have been scaled from a small dataset analysed on a SUN
Ultra II. Two further considerations immediately apply.
\begin{itemize}
\item Moving to parallel architectures will give significant reduction
in these timings. Implementation of each algorithm on the 512
processor Cray T3E at NERSC indicates that the improvement can be up
to a factor of 1000. However, this does assume that we continue to
keep all the necessary matrices simultaneously in core; any reduction
to vector operations, relocation to disc, or recalculation will
dramatically reduce this improvement.
\item The datasets under consideration will be obtained incrementally
over the next 10 years. We should therefore take into consideration
Moore's law --- that computer power doubles every 18 months --- to
allow for corresponding increases in available memory and speed.
Current trends do not, however, suggest any significant increase in
the total parallel processor time ($O(10^{4})$ hours) available to us.
\end{itemize}
Taken together, we can conclude that these algorithms, judiciously
applied, will be sufficient to analyse $10^{4}$ pixel datasets
immediately, the $10^{5}$ pixel datasets expected in the next 2 years
some 6 years from now, and the $10^{6}$ pixel datasets expected in 5
-- 10 years only 16 years from now. However, since we would like to be
able to analyse not only the actual datasets as soon as they are
obtained, but also simulated datasets in advance of the observations,
improved algorithms are still essential.

\section*{Acknowledgments}

This work was supported by the Laboratory Directed Research and
Development Program of Lawrence Berkeley National Laboratory under the
U.S. Department of Energy, Contract No. DE-ACO3-76SF00098, and used
resources of the National Energy Research Scientific Computing Center,
which is supported by the Office of Energy Research of the
U.S. Department of Energy. This work is also part of the COMBAT
project supported by NASA AISRP grant NRA-96-10-OSS-047. The author
wishes to thank Andrew Jaffe for many enlightening discussions.

\newpage

\begin{center}
\begin{tabular}{|cc|c|c|}
\hline & & & \\ [-0.1in] 
\multicolumn{2}{|c|}{TERM} & MEMORY & OPERATIONS \\ [0.05in] 
\hline 
& & & \\ [-0.1in] 
$X_{b} = M^{-1} \frac{\partial S}{\partial C_{b}}$ & $\forall \; b$ & $O(N_{b} \, N_{p}^2)$ & $O(N_{
b} \, N_{p}^3)$ \\ [0.05in] 
\hline 
& & & \\ [-0.1in]
${\rm Tr} \left[ X_{b} X_{b'} \right]$ & $\forall \; b, \, b'$ & $O(N_{p}^2)$ & $O(N_{b}^{2} \, N_{p
}^2)$ \\ [0.05in]
\hline
& & & \\ [-0.1in] 
$X_{l} = M^{-1} Y_{l}$ & $\forall \; l$ & $O(N_{l}^{2} N_{p})$ & $O(N_{l}^{2} \, N_{p}^2)$ \\ [0.05in] 
\hline
& & & \\ [-0.1in] 
$Z_{l \, l'} = Y_{l}^{T} X_{l'}$ & $\forall \; l, \, l'$ & $O(N_{l}^{4})$ & $O(N_{l}^{4} \, N_{p})$ \\ [0.05in] 
\hline 
& & & \\ [-0.1in] 
${\rm Tr} \left[ Z_{l \, l'} Z_{l \, l'}^{T} \right]$ & $\forall \; l, \, l'$ & $O(N_{l}^{2})$ & $O(N_{l}^{4})$ \\ [0.05in] 
\hline
\end{tabular}
\\ [0.2 in] Table 1: Scaling in the calculation of the Fisher matrix
$F$ for the two quadratic estimator algorithms A1 (first two rows), A2
(last three rows).
\end{center}

\newpage

\begin{center}
\begin{tabular}{|c|c|c|c|c|c|c|c|}
\hline
& & & & \multicolumn{2}{c|}{\mbox{}} & \multicolumn{2}{c|}{\mbox{}} \\ [-0.1in]
& & & & \multicolumn{2}{c|}{SIZE} & \multicolumn{2}{c|}{TIME} \\ [0.1in]
\cline{5-8}
& & & & & & & \\ [-0.1in]
DATASET & $N_{p}$ & $N_{b}$ & $N_{l}$ & A1 & A2 & A1 & A2 \\ [0.1in]
& & & & $O(N_{b} \, N_{p}^2)$ & $O(N_{l}^{4})$ & $O(N_{b} \, N_{p}^3)$ & $O(N_{l}^{4} \, N_{p})$ \\ [0.1in]
\hline
& & & & & & & \\ [-0.1in]
COBE & $10^{3}$ & 30 & 30 & 240 Mb & 8 Mb & 15 min & 1 min \\ [0.1in]
\hline
& & & & & & & \\ [-0.1in]
MAXIMA/ & $10^{4}$ & 20 & 1000 & 16 Gb & 8 Tb & 7 days & 20 years \\
BOOMERANG & to $10^{5}$ & 20 & 1000 & 1.6 Tb & 8 Tb & 20 years & 200 years \\ [0.1in]
\hline
& & & & & & & \\ [-0.1in]
MAP/PLANCK & $10^{6}$ & 1000 & 1000 & 8 Pb & 8 Tb & 1 Myears & 2 Kyears \\ [0.1in]
\hline
\end{tabular}
\\ [0.2 in] 
Table 2: Size and time costs for the calculation of the Fisher matrix
$F$ for archetypal datasets on a SUN Ultra II for the two quadratic
estimator algorithms A1, A2.
\end{center}


\frenchspacing

\begin{thebibliography}{99}

\bibitem{S} G. Smoot, astro-ph/9705135

\bibitem{Kn} L. Knox, {\em Phys. Rev.} {\bf D52}, 4307 (1995),
astro-ph/9504054

\bibitem{HSS} W. Hu, N. Sugiyama and J. Silk, {\em Nature} {\bf 386},
37, (1997), astro-ph/9604166

\bibitem{KKJS} A. Kosowsky, M. Kamiankowski, C. Jungman and D. Spergel
{\em Nucl. Phys. Proc. Suppl.} {\bf 51B}, 49, (1996), astro-ph/9605147

\bibitem{G1} K. M. G\'{o}rski, {\em Ap. J.} {\bf 430}, L85, (1994),
astro-ph/9403066

\bibitem{G2} K. M. G\'{o}rski, G. Hinshaw, A. J. Banday, C. L. Bennett, 
E. L. Wright, A. Kogut, G. F. Smoot and P. Lubin {\em Ap. J.} {\bf
430}, L89, (1994), astro-ph/9403067

\bibitem{BJK} J. R. Bond, A. H. Jaffe and L. Knox, (submitted to {\em
Phys. Rev.} {\bf D}) astro-ph/9708203.

\bibitem{Ko} A. Kogut, presented at the INPAC/ITP Conference on `CMB
Data Analysis and Parameter Extraction', Santa Barbara (November
1997).

\bibitem{OSH} S. P. Oh, D. N. Spergel and G. Hinshaw, (in preparation)
(1997).

\bibitem{T} M. Tegmark, {\em Phys. Rev.} {\bf D55} 5895 (1997),
astro-ph/9611174.

\end{thebibliography}
\end{document}